Extreme Solar Eruptions and their Space Weather Consequences

Nat Gopalswamy

NASA Goddard Space Flight Center, Greenbelt, MD 20771, USA

Abstract:

Solar eruptions generally refer to coronal mass ejections (CMEs) and flares. Both are important sources of space weather. Solar flares cause sudden change in the ionization level in the ionosphere. CMEs cause solar energetic particle (SEP) events and geomagnetic storms. A flare with unusually high intensity and/or a CME with extremely high energy can be thought of examples of extreme events on the Sun. These events can also lead to extreme SEP events and/or geomagnetic storms. Ultimately, the energy that powers CMEs and flares are stored in magnetic regions on the Sun, known as active regions. Active regions with extraordinary size and magnetic field have the potential to produce extreme events. Based on current data sets, we estimate the sizes of one-in-hundred and one-in-thousand year events as an indicator of the extremeness of the events. We consider both the extremeness in the source of eruptions and in the consequences. We then compare the estimated 100-year and 1000-year sizes with the sizes of historical extreme events measured or inferred.

## 1. Introduction

Human society experienced the impact of extreme solar eruptions that occurred on October 28 and 29 in 2003, known as the Halloween 2003 storms. Soon after the occurrence of the associated solar flares and coronal mass ejections (CMEs) at the Sun, people were expecting severe impact on Earth's space environment and took appropriate actions to safeguard technological systems in space and on the ground. The high magnetic field in the CMEs indeed interacted with Earth's magnetic field and produced two super intense geomagnetic storms. Both CMEs were driving strong shocks that accelerated coronal particles to GeV energies. The shocks arrived at Earth in less than 19 hours. The consequences were severe: in Malmoe, a southern city in Sweden, about 50,000 people experienced a blackout when the transformer oil heated up by 10ºC. About 59% of the reporting spacecraft and about 18% of the onboard instrument groups were affected by these events. In order to protect Earth-orbiting spacecraft from particle radiation, they were put into safe mode (Webb and Allen, 2004). The high energy particles from the CMEs penetrated Earth's atmosphere causing significant depletion of stratospheric ozone. The ionospheric total electron content over the US mainland increased tenfold during 30–31 October 2003. Significant enhancement of the density in the magnetosphere also coincided with the arrival of the CMEs at Earth. In addition to the Earth's space environment, the impact of the CMEs was felt throughout the heliosphere, all the way to the termination shock. The detection of the impact was possible because there were space missions located near Mars (Mars Odyssey), Jupiter (Ulysses), and Saturn (Cassini) as well as at the outer edge of the solar system (Voyager 1 and 2). The MARIE instrument on board the Mars Odyssey mission was completely damaged by the energetic particles from these CMEs. The widespread impact of these Halloween events have been documented in about seventy articles published during 2004-2005 (see Gopalswamy et al. 2005a for the list of the articles). The solar active region from which the CMEs originated also was very large and had the potential to launch energetic CMEs.

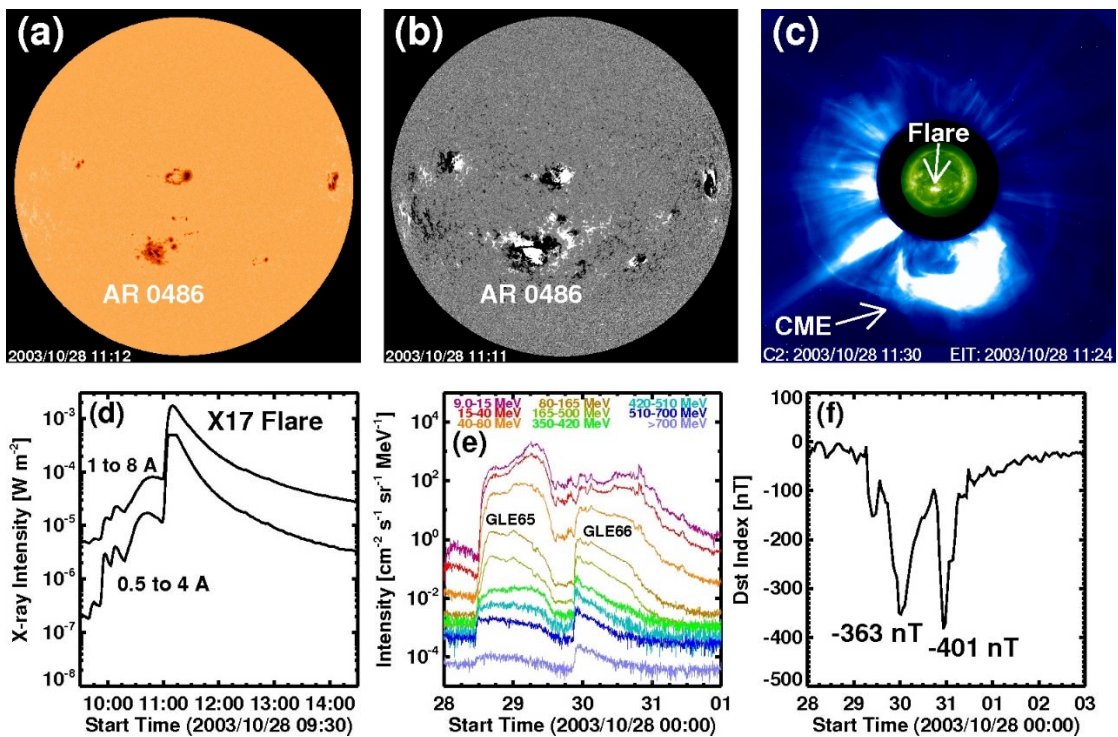

Figure 1. The solar source and space weather consequences of the 2003 October 28 CME. (a) a continuum image of the Sun from SOHO/MDI showing the sunspot region 10486 (Sunspots appear dark because they are ~ 2000 K cooler than the surrounding photosphere at ~6000 K).; (b) the sunspot region as seen in a SOHO/MDI magnetogram (white is positive and black is negative magnetic field region); (c) A SOHO/LASCO white-light image with superposed SOHO/EIT

image showing the flare brightening from the active region 10486 (the dark disk is the occulting disk); (d) GOES soft X-ray light curve showing the X17 flare in two energy channels (1to 8 Å and 0.5 to 4 Å) , (e) GOES proton intensity in various channels, including the >700 MeV channel indicative of ground level enhancement (GLE) associated with the eruption (GLE65) in (c) as well as the next one (GLE66); (f) Dst index from World Data Center, Kyoto showing the superstorms with Dst = -363 nT associated with the eruption in (c) and Dst = -401 nT associated with the next eruption on 2003 October 29. This figure illustrates the chain of events from the Sun to Earth's magnetosphere considered throughout this paper: active regions, flares, CMEs, SEP events, and geomagnetic storms.

Figure 1 shows the source active region (10486) with sunspots and its complex magnetic structure as observed by the Magnetic and Doppler Imager (MDI) on board the Solar and Heliospheric Observatory (SOHO). The region produced two of the Halloween events that are of historical importance. The first eruption on 28 October 2003 was seen bright in EUV wavelengths and had the soft X-ray flare size of X17. The CME was a symmetric halo as seen by the Large Angle and Spectrometric Coronagraph (LASCO) on board SOHO. The 28 and 29 October 2003 eruptions were responsible for intense SEP events that had ground level enhancements (GLEs) numbered GLE 65 and GLE 66, respectively. The CMEs had speeds exceeding 2000 km/s and produced super magnetic storms (Dst < -200 nT) when they arrived at Earth. The Halloween solar eruptions thus turned out to be extreme events both in terms of their origin at the Sun and their consequences in the heliosphere.  The two events were observed extremely well by many different instruments from space and ground and the knowledge on space weather events helped us to take appropriate actions to limit the impact when possible. The Sun must have produced such events many times during its long history of 4.5 billion years, but the occurrence now has high significance because the human society has become increasingly dependent on technology that can be affected by solar eruptions. It is of interest to know the origin of the extreme events and how big an impact they can cause.

An overview of extreme events on the Sun and their heliospheric consequences is provided in section 2.  Extreme event sizes are estimated in section 3 for CMEs, flares, and source active regions assuming the extreme events to be located on the tails of various cumulative distributions. Section 4 considers the heliospheric response of solar eruptions in the form of SEP events and geomagnetic storms. The chapter is summarized in section 5.

## 2. Overview of Extreme Events

The definition of an extreme event is not very concrete, but can be thought of as an event on the tail of a distribution. An extreme event can also be thought of as an occurrence that has unique characteristics in its origin and/or in its consequences. For example, a CME that has an extreme speed can be considered as an extreme event if such an occurrence is extremely rare. Among the thousands of CMEs observed by the Solar and Heliospheric Observatory (SOHO) from 1996 to 2015, only a couple have speeds exceeding 3000 km/s Therefore, one can consider a CME with speed exceeding 3000 km/s as an extreme event. But how high can the CME speed get? To answer this question, one has to consider the energy source of CMEs and how that energy is converted to CME kinetic energy. It has been established that CMEs can only be powered by the magnetic energy in closed magnetic field regions on the Sun (see e.g., Forbes, 2000). There are two types of closed field regions that are known to produce CMEs: sunspot regions (active regions) and quiescent filament regions (see e.g., Gopalswamy et al. 2010). Observations have shown that the fastest CMEs originate from active regions because they possess the high magnetic energy needed to power such CMEs.  The magnetic energy of an active region depends on its size and the average field strength. Historically, there is a long record of sunspot area, which can be taken as a measure of the active region area. The magnetic fields in sunspots were discovered in 1908 by George Ellery Hale (Hale, 1908) and have been recorded since then with routine field measurements starting in 1915. Following the work of Mackay et al. (1997), one can compute the potential energy in active regions as a measure of the maximum free energy available to power eruptions (see e.g., Gopalswamy 2011). Essentially, this procedure traces the origin of extreme CMEs to the extremeness in the source region of CMEs, although additional considerations such as the conversion efficiency from the magnetic energy to CME kinetic energy play a role.

Another manifestation of a solar eruption is the flare, which is primarily identified with the sudden increase in electromagnetic emission from the Sun at various wavelengths. The flare phenomenon was originally discovered in white light by Carrington (1859) and Hodgson (1859) and has been extensively observed in the H-alpha line since the beginning of the 20[th] century.

The most common way of flare detection at present is in soft X-rays and the flare size is indicated by the intensity expressed in units of W m$^{-2}$ in the 1-8 Å channel. Flares of size $10^{-4}$ W m$^{-2}$ are classified as X-class. The largest flare ever observed in the space age had an intensity of X28 or $2.8\times10^{-3}$ W m$^{-2}$ observed on 2003 November 4 from the same active region 10486 a few days after the eruptions described in Fig. 1. The flare was accompanied by a fast (~2700 km/s) CME with a kinetic energy of ~$6\times10^{32}$ erg (Gopalswamy et al. 2005b).

The primary consequences of CMEs are large SEP events and geomagnetic storms, both of which are sources of severe space weather (see e.g., Gopalswamy 2009a). Corotating interaction regions can also cause geomagnetic storms that are more frequent but less severe compared to CMEs (see e.g., Borovsky and Denton, 2006). We do not consider them here. The particles in large SEP events are accelerated at the CME-driven shock, while geomagnetic storms depend on the CME speed and its magnetic content. Each of these space weather events has a chain of effects on Earth's magnetosphere, ionosphere, atmosphere, and even on the ground. In addition, SEPs pose radiation hazard to astronauts and adversely affect space technology in the near-Earth as well as interplanetary space. It must be noted that SEPs are accelerated also at the flare site, which are responsible for a different types of electromagnetic emission when they propagate toward, and interact with, the solar surface. However, their contribution to the observed SEPs in space is not fully understood, but is usually small compared to that from CME-driven shocks (see e.g., Reames 2015). Some studies suggest that flares are the dominant sources of high-energy SEPs observed in the interplanetary medium (see e.g., Dierckxsens, 2015; Grechnev et al., 2015; Trottet et al., 2015). Cliver (2016) points out that the conclusion is not supported if all the SEP events are included in the correlative analyses. Particles are energized by other mechanisms throughout the heliosphere, providing seed particles to the shock acceleration process (see e.g., Mason et al. 2013; Zank et al. 2014). We considered recent studies (Dierckxsens et al. 2015; Grechnev et al. 2015; Trottet et al. 2015) that suggest that solar flares are significant sources of the high-energy protons observed in interplanetary space following solar eruptions and may, in fact, be the dominant accelerator of such protons.

The electromagnetic emission from solar flares generally cause excess ionization in the ionosphere, thereby changing the ionospheric conductivity. For example solar-flare X-rays cause sudden ionospheric disturbances that can affect radio communications. Intense radio bursts are produced by energetic electrons accelerated during flares. If the frequencies of the radio bursts are close to those of GPS and radar signals, the bursts can drown the signals out (see e.g., Kintner et al. 2009).

The extreme space weather consequences thus depend on extreme CME properties. In the case of SEP events, one can think of very strong shocks, which ultimately result from very high CME speeds. Geomagnetic storms also depend on CME speeds as they arrive at Earth's magnetosphere, but they also require intense southward magnetic field in the CME and/or in the shock sheath (e.g., Wu and Lepping 2002; Gopalswamy et al. 2008; 2015a). The storm strength (as measured by say, the Dst index) can also depend on solar wind density, but the effect is not significant for extreme storms we are interested in (e.g., Weigel 2010). High-speed shocks that arrive at Earth in less than a day are known as fast transit events (Cliver et al. 1990; Gopalswamy et al. 2005b). These shocks are considered to be extreme events because they can cause high levels of energetic storm particles (ESPs) at Earth and compress the magnetosphere observed as sudden impulse or sudden commencement (SC) of geomagnetic storms (Araki, 2014). Such shocks are also very strong near the Sun and are highly likely to accelerate SEPs to very high energies. The resulting SEP spectrum is expected to be hard leading to high-energy particles that affect the Earth's ionosphere and atmosphere.

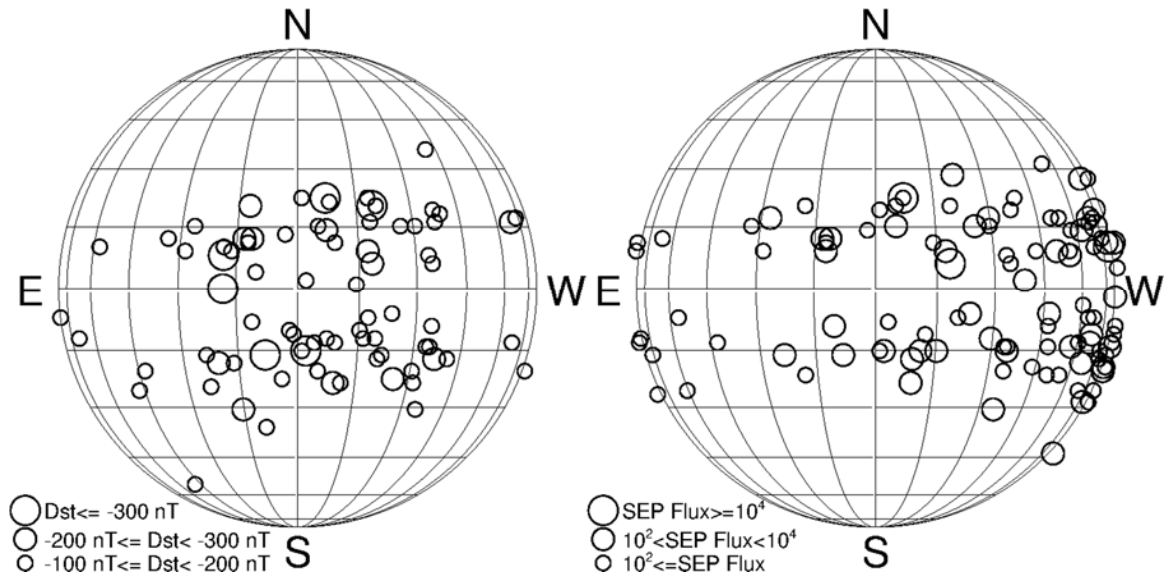

Figure. 2. Solar sources of CMEs causing intense geomagnetic storms (Dst ≤ -100 nT) (left) and large SEP events (intensity >10 pfu in the >10 MeV channel; pfu is the particle flux unit defined

as 1 pfu = 1 particle per (cm$^2$ s sr)) (right) during 1996 to 2016. The size of the circle indicates the intensity of the event as noted on the plots. The latitude and longitude grids are 15º apart. No correction was made for the solar B0 angle, the heliographic latitude of the central point on the solar disk (updated from Gopalswamy 2010a).

The consequences at Earth become extreme only under certain conditions because Earth presents only a small cross section to solar events. This is illustrated in Fig. 2 as the distribution of solar sources of CMEs that caused intense geomagnetic storms (Dst ≤ -100 nT) and large SEP events (>10 MeV proton intensity >10 pfu).  The size of the circles denote the intensity of events. The most intense geomagnetic storms are associated with CMEs originating very close to the solar disk center (no Dst ≤ -300 nT events beyond a central meridian distance (CMD) of ~20º). Beyond a CMD of ~30º, we see only the weaker storms. CMEs originating from close to the disk center head directly toward Earth and deliver a head-on blow to Earth's magnetosphere. This fact was established long ago by Hale (1931) and Newton (1943). The source regions of CMEs producing SEP events have a different distribution: the most intense SEP events generally originate from the western hemisphere of the Sun. At CMD >30º in the eastern hemisphere, SEP events are less frequent and weak (peak >10 MeV intensity <100 pfu). The reason is the heliospheric magnetic structure, which takes the form of an Archimedean spiral (Parker Spiral) along which accelerated particles propagate.  Magnetic field lines originating from the western hemisphere of the Sun are connected to Earth and hence particles can be detected at Earth. Thus an extremely fast CME from the east limb may not produce an extreme space weather event at Earth. However, planets or spacecraft located above the east limb could be affected by such CMEs. There is no such source restriction for solar flares: electromagnetic emissions from flares reach Earth so long as they occur on the frontside of the Sun.

We are interested in extreme events both in their origins at the Sun and their consequences and space weather effects. At the Sun, we are interested in the size and magnetic field in active regions as well as the amount of energy that can be stored and released in them. The immediate consequences of the energy release are flares and CMEs. CMEs drive shocks that accelerate SEPs from near the Sun and into the heliosphere; they also cause sudden commencement when arriving at Earth. CMEs cause severe geomagnetic storms when they have appropriate field orientation, field strength, and speed. We use cumulative distributions of these events, fit a function to the tail of the distributions, and estimate the size of a one-in-100- and one-in-1000-year events. Traditionally the power-law distribution has been extensively used (e.g., Nita et al. 2002; Song et al. 2012; Riley 2012), which can lead to overestimates for some types of events. Other distributions such as a lognormal distribution have been found to better represent the data and provide better confidence intervals for extreme-event estimation (Love et al. 2015). Here we use both a power law (e.g., Clauset et al. 2009; Aschwanden et al. 2016) and a version of the Weibull distribution (Weibull, 1951). Our main is to extend the tail to smaller probability regimes without worrying about the theoretical basis of the distributions. Such an approach seems to be consistent with some of the historical extreme events, but may not be unique. It should also be made clear that inferring events on the tail of distributions assumes that the same physics is involved in the inferred parametric regime.

## 3. Estimates of Extreme Events

### 3.1 CME speeds

Since CMEs are the most energetic phenomena relevant to space weather, we start with the extreme CME events. One of the basic attributes of CMEs is their speed in the coronagraph field of view (FOV). CMEs start from zero speed during eruption, attain a peak speed and then tend to slow down. Observations close to the Sun that were occasionally available in the early phase of SOHO mission (Gopalswamy and Thompson 2000; Zhang et al. 2001; Cliver et al. 2004),the STEREO mission (see e.g., Gopalswamy et al. 2009a; Bein et al. 2011), and the ground based Mauna Loa Coronameter (St Cyr t al. 2015; Gopalswamy et al. 2012) have shown that CMEs attain a peak acceleration ranging from a fraction to several km s$^{-2}$ near the Sun (typically at heliocentric distances <3 Rs). Once the acceleration ceases, CMEs move with constant speed or slowly decelerate due to the drag force exerted by the ambient medium. The average speed in the coronagraph FOV (2-32 Rs) ranges from ~100 km/s to >3000 km/s. CMEs causing space weather typically have higher speeds.

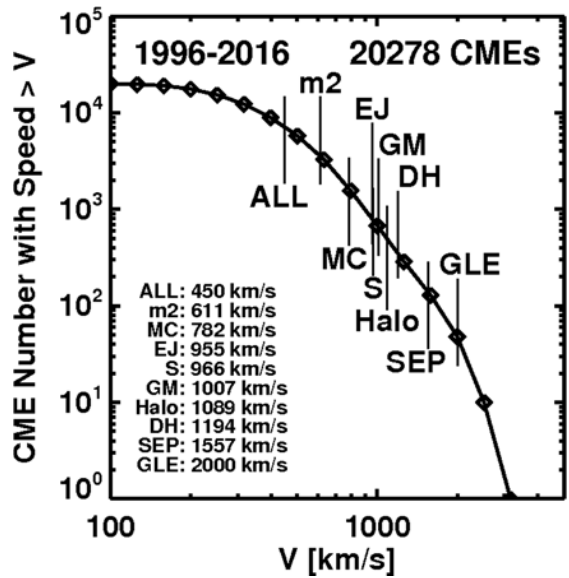

Figure 3. Cumulative distribution of CME speed (V) from SOHO/LASCO. The CME speeds from https://cdaw.gsfc.nasa.gov have been measured in the sky plane and no corrections have been applied. The average speeds of CME populations responsible for various coronal and interplanetary phenomena are marked on the plot. The 10 November 2004 CME at 02:26 UT had the highest speed of 3387 km s$^{-1}$. Updated from Gopalswamy (2016).

## 3.2 Distribution Functions for CME Speeds and Kinetic Energies

Figure 3 shows the cumulative distribution of CME speeds measured in the FOV of SOHO's Large Angle and Spectrometric Coronagraph (LASCO). The average speeds of several CME populations responsible for energetic phenomena are noted on the plot. The CMEs are related to: metric type II radio bursts (m2) due to shocks in the corona at heliocentric distances <2.5 Rs; magnetic clouds (MC), which are the inter planetary CMEs (ICMEs) with flux rope structure; ICMEs lacking flux rope structure and hence named as ejecta (EJ); interplanetary shocks (S) detected in the solar wind; geomagnetic storms (GM) caused by CME magnetic field or shock sheath; halo CMEs (Halo) that appear to surround the occulting disk of the coronagraph and propagating Earthward or anti-Earthward; decameter-hectometric (DH) type II bursts indicating electron acceleration by CME-driven shocks in the interplanetary medium; SEP events caused by CME-driven shocks; ground level enhancement (GLE) in SEP events indicating the acceleration of GeV particles. The average speed of every one of these populations is significantly greater than the average speed of the general population (450 km/s). It must be noted that MC, EJ, GM, and Halo are related to the internal structure of CMEs in the solar wind, while the remaining are all related to the shock-driving capability of CMEs. All these CME populations are generally related. What is remarkable about the cumulative distribution is that there are not many CMEs with speeds exceeding ~3000 km/s. Gopalswamy et al. (2010) attributed the lack of CMEs to speeds >3000 km/s to the free energy that can be stored in solar active regions and the fractionation of the released energy in the form of CMEs.

The fastest CME in Fig. 3 occurred on 2004 November 10 at 02:26 UT. The average speed in the coronagraph FOV was 3387 km/s. One might wonder if the high speed of the CME was because of the preceding CMEs that sweeps out the ambient material, presenting a low-density (and hence low-drag) medium to the succeeding CME. But the drag depends not only on density, but also on the CME surface area and the square of the excess speed of the CME over the ambient medium. When a low-density medium is created, a CME propagating through such a medium expands and hence acquires a greater area that increases the drag. Similarly, the high speed also increases the drag. So, the net effect may not be a significant decrease in drag. In fact, the 2004 November 10 CME was observed to slow down within the coronagraph FOV (https://cdaw.gsfc.nasa.gov/CME_list/UNIVERSAL/2004_11/htpng/20041110.022605.p302s.htp.html), even though there was a preceding CME from the same source region nine hours before. Therefore, the initial high speed is likely to be due to the propelling force (solar source property) rather than the drag force (ambient medium property).

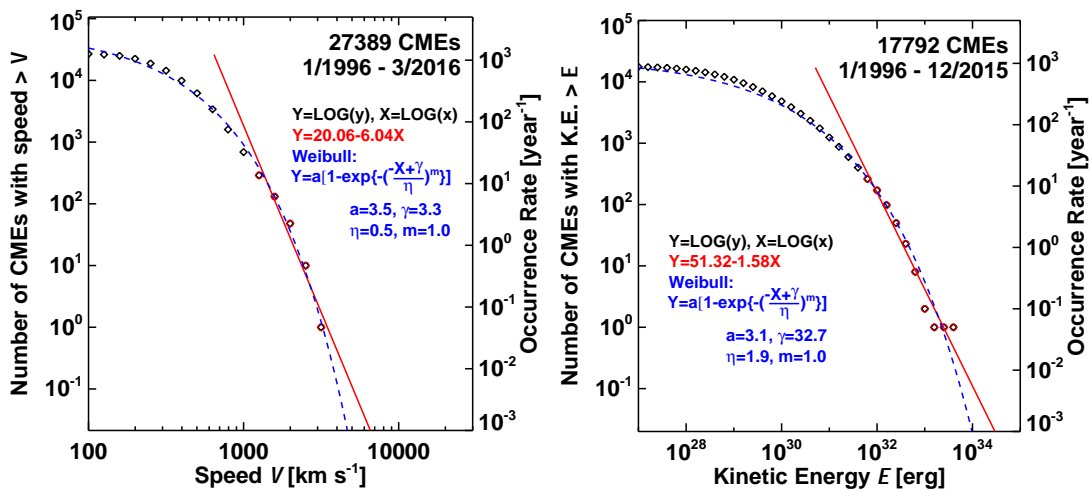

Figure 4. Cumulative distribution of CME speeds (left) and kinetic energies (right) from SOHO/LASCO catalog (https://cdaw.gsfcs.nasa.gov) for the period 1996-2016. Power-law (e.g.,

Clauset et al. 2009) and Weibull (Weibull 1951) fits to the data points are shown. The speed and kinetic energy data points are obtained by binning the original data into 5 data points per decade. The 10 November 2004 CME at 02:26 UT has the highest speed of 3387 km s$^{-1}$ and the 9 September 2005 CME at 19:48 UT has the highest kinetic energy of $4.20 \times 10^{33}$ erg.

Figure 4 shows the cumulative distributions of speeds and kinetic energies of CMEs observed over the past two decades. We have used a power law and the Weibull functions to fit the data points. Clearly the power law is applicable only over a very limited range of speeds and kinetic energies. On the other hand, the Weibull function fits much better over the entire range, although it has more free parameters. The steep drop in the number of events at high speeds and kinetic energies seems to be real because with current cadence of the LASCO is high enough that energetic CMEs are not missed. An event on the tail of the Weibull distribution in Fig. 4 may occur once in 100 years with a speed of 3800 km/s while a once in thousand year event will have a speed of ~4700 km/s. We refer to these events as once-in-100-year and once-in-1000-year events to denote the event size expected once in 100 years and once in 1000 years, respectively. Hereafter we refer to these events as 100-year and 1000-year events for simplicity. From Fig. 4, we can infer that the 100-year and 1000-year kinetic energies as $4.4 \times 10^{33}$ and $9.8 \times 10^{33}$ erg, respectively. It must be noted that these kinetic energies are only a few times greater than the highest reported values. We shall return to the reason for these limiting values later.

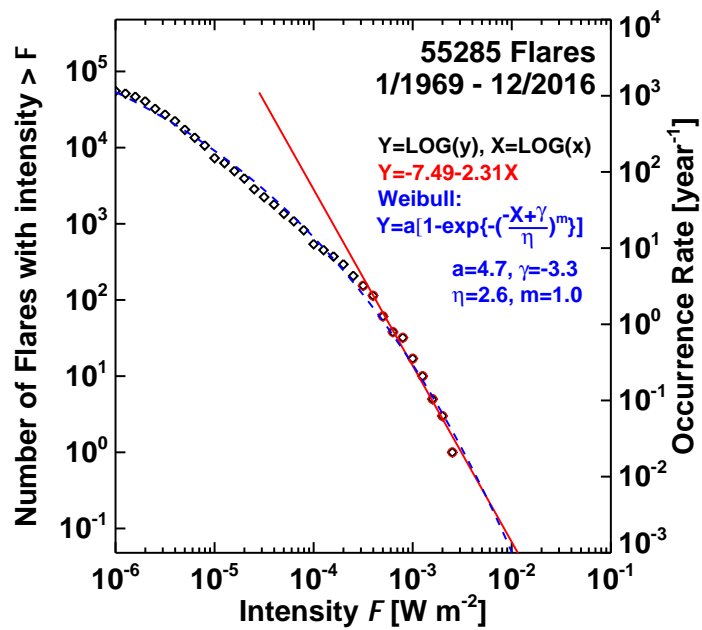

Figure 5. Cumulative distribution of flare sizes between 1969 and 2016. Weibull and power-law fits to the data points are shown. The 4 November 2003 flare at 19:29 UT has the highest intensity of $2.8 \times 10^{-3}$ W m$^{-2}$ (X28). The flare data are from https://www.ngdc.noaa.gov/stp/solar/solarflares.html.

### 3.3 Flare Size Distribution

Solar flares typically accompany CMEs, but many also occur without CMEs. CMEless flares are confined typically have an upper limit to their sizes: ~X2.0. About 10% of X-class flares are known to lack CMEs (Gopalswamy et al. 2009b). Here we consider the cumulative distribution of all the flares that have been recorded by various GOEs satellites since 1969 in the 1-8Å energy band (see Fig. 5). The distribution shows a break around the X2 level ($2 \times 10^{-4}$ W m$^{-2}$). According to the Weibull distribution, the 100-year and 1000-year event sizes are X43.9 and X101 respectively. The power law distribution yields similar flare sizes: X42 and X115. The 100-year size is similar to the estimated size of the 2003 November 4 soft X-ray flare ever recorded in the 1-8 Å energy band by the GOES satellites (Woods et al. 2004; Thomson et al. 2004; Brodrick et al. 2005). The data point corresponding the largest flare size (X28) in Fig. 5 represents this event. It must be noted that the GOES X-ray sensor saturated at a level of X17.4 for about 12 minutes, so the X28 value was an initial estimate. Brodrick et al. (2005) concluded that the flare size should be in the range X34–X48, with a mean value of X40. The corrected data point is close to the fitted lines corresponding to the Weibull and power-law functions. Based on solar flare effects on the ionosphere, it has been concluded that the 1859 September 1 flare should have been at least as strong as the 2003 November 4 flare. The flare size estimate for the Carrington flare is in the range X42 – X48, with a nominal value of X45 (see Cliver and Dietrich 2013 and references therein). It is remarkable that the Weibull distribution provides an estimate consistent with several independent estimates of the peak values of the 2003 November 4 flare and the Carrington flare.

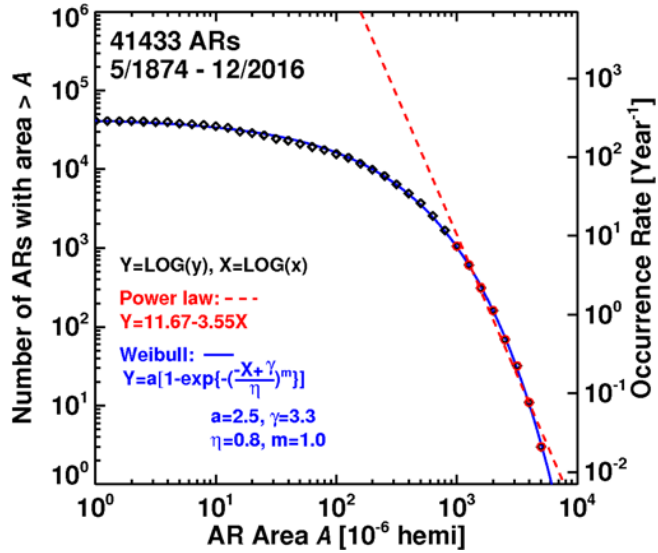

Figure 6. Cumulative number of active region (sunspot group) areas *A* from 1874 to 2016. *A* is expressed in microhemispheres (also known as millionths of solar hemisphere, msh; 1 msh = $3.07 \times 10^{16}$ cm$^2$). On the right hand side Y-axis, the occurrence rate per year (number of active regions in each bin divided by the data interval of 143.5 years). Sunspot group areas are derived from daily photographic images of the Sun recorded at the Royal Greenwich Observatory (ftp://ftp.ngdc.noaa.gov/STP/SOLAR_DATA/SUNSPOT_REGIONS/Greenwich/) for the period 1874–1976. The area data have been extended beyond 1976 by the Solar Observing Optical Network (SOON, https://www.ngdc.noaa.gov/stp/space-weather/solar-data/solar-imagery/photosphere/sunspot-drawings/soon/).

The size of the 1000-year flare (X101) is only a factor of ~2 larger than the Carrington flare. The bolometric energy corresponding to an X100 flare is $10^{33}$ erg (see e.g., Benz et al. 2008). Flares with bolometric energies >$10^{33}$ erg are considered as super flares (Schafer et al. 2000; Maehara et al. 2012; Shibata et al. 2013). Thus, the tail of the flare-size distribution suggests that super flares can occur on the Sun once in a millennium. A $10^{34}$ erg flare can occur on the Sun only once in 125,000 years, too infrequent compared to the once-in-800-years occurrence suggested by Shibata et al. (2013).

### 3.4 Active Regions and their Magnetic Fields

One of the important parameters related to the origin of solar eruptions is the sunspot area, which has been known for a long time. We refer to the sunspot group area as the active region area. We use the whole sunspot area, which includes the penumbra, not just the umbra.  Figure 6 shows the cumulative distribution of the active region area $A$ for the period 1874 to 2016. The cumulative number decreases slowly until the area reaches ~1000 msh (millionths of solar hemisphere) and then decreases rapidly. The Y-axis on the right hand side gives the occurrence rate N per year. The rapidly declining part of the distribution fits to a power law, N [yr$^{-1}$] = $4.68 \times 10^{11} A^{-3.55}$. The maximum observed area was ~5000 msh. Such large-area active regions were observed only twice over the 143 year period used in the distribution (both of these active regions occurred in solar cycle 18). All the data points can also be fit to the Weibull's function, which agrees with the power law at high $A$. Note that we modified the Weibull function by introducing an additional scale factor '$a$'. According to the power law, a 100-year active region has an area of ~7000 msh and will be considered as an extreme event. The Weibull function gives a slightly lower area for the 100-year active region: ~5900 msh.  The observation that superflares tend to occur in solar-like stars with large spot areas (more than order of magnitude larger than the largest sunspot areas) suggest that the same physical process is responsible for the formation of active regions in the Sun and other solar-like stars (Maehara et al. 2017).

While the active region area has been measured systematically since the late 1800s, the measurement of sunspot magnetic fields started only around 1915. There have been a number of investigations in the 20[th] century that found a good correlation between the sunspot area and the maximum field strength in the umbra (Livingston et al. 2006 and references therein). These investigations also found that the number of regions with field strengths >5000 G is exceedingly low.  Livingston et al (2006) compiled sunspot field measurements of 12804 active regions in the interval 1917-2004. The cumulative distribution shows only five active regions with sunspot field strengths >5000 G, one of them being 6100 G. These authors also noted that the distribution was a steep power law with an index of -9.5. The relation between sunspot field strength $B$ (G) and the active region area $A$ (msh) has been found to be of the form (Ringnes and Jensen, 1960; Nagovitsyn et al. 2017):

$B = p \log A + q$   (1)

where $p$ and $q$ are coefficients that seem to vary between solar cycles and $B$ is in units of 100 G. Ringnes and Jensen (1960) reported that $p$ and $q$ varied significantly between cycles. For a particular period (1945 – 1948) with the highest correlation (r = 0.92 for 43 sunspots) between $B$ and log$A$, $p$ = 23.3 and $q$ = -27.0.  Nagovitsyn et al. (2017) confirmed this B – A relationship for cycles 23 and 24, although they found only $q$ varied between the two cycles. It must be note that Norton et al. (2013) considers a shorter period (3 years in the declining phase of cycle 24) and does not find a significant variation in the umbral field strength. Schad (2014) finds a non-linear relationship between magnetic field strength and umbral size, over a short period (6 years).  Note that both Norton et al. (2013) and Schad (2014) consider umbral areas, not the whole area including the penumbra.

Given the good correlation between the sunspot area and sunspot $B$, we see that the sharp decline in the number of events with $B$ is consistent with the rapid drop in the number of events with large active region area. Thus an active region with $A \sim 6000$ msh is expected to have a $B$ of ~6100 G, very similar to the extreme case reported by Livingston et al. (2006). It must be noted that the peak field strength is found in the umbra of sunspots, but not throughout the active region area. Nevertheless, we can consider a hypothetical active region with an area of ~6000 msh and a peak field strength of 6100 G as an extreme case of source active region that will be used for further discussion. The maximum possible magnetic potential energy (MPE) can be computed as $(B^2/8\pi)A^{1.5}$, where $B$ is the magnetic field strength of the active region that has a sunspot area $A$. For $B$ = 6100 G and $A$ = 6000 msh, we get MPE = $3.7 \times 10^{36}$ erg.

The sunspot magnetic field is thought to emerge from the toroidal field located at the base of the convection zone in the solar interior (see e.g., Basu 2016). Based on helioseismic techniques, Basu (1997) and Antia et al. (2000) have estimated an upper limit of $3 \times 10^5$ G for the field strength at the base of the convection zone. The field strength measured on the surface is about two orders of magnitude smaller than the one at the base of the convection zone.  The limit of the field strength in the solar interior ultimately seems to be the physical reason for the size and field strength in solar active regions that determine the free energy available to power eruptions.

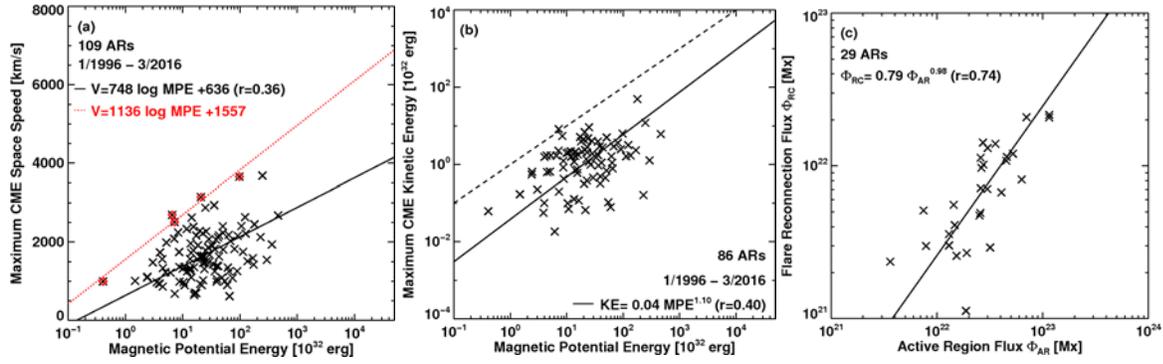

Figure 7. Scatter plot of the magnetic potential energy (MPE) of active regions with maximum speed (a) and maximum kinetic energy (KE) of CMEs from the active regions (b). Only CMEs with speeds > 500 km/s are included. The regression lines are shown in solid black. In the speed plot, the red line fits the top 5 data points. In the kinetic energy plot, the dashed line represents equal energies. The correlation coefficients r = 0.36 and r = 0.40 are significant despite the large scatter because the corresponding Pearson's critical values are 0.316 and 0. 349, respectively, for a significance level of 99.95%. Note that we used deprojected speeds ($V_{sp}$) as opposed to sky-plane speeds ($V_{sky}$) used in Figs. 3 and 4. Speeds of full halo CMEs and partial halos are deprojected using a cone model or the empirical formula $V_{sp} = 1.10V_{sky} + 156$ km/s (Gopalswamy et al. 2015b) when CMD < 60º. For CMEs with CMD > 60º a simple geometrical deprojection used. (c) A scatter plot between the reconnected (RC) flux during an eruption and the total flux in the source active region indicates a good correlation (r = 0.74 with a Pearson critical correlation coefficient of 0.579 at 99.95% confidence level for 29 active regions). RC fluxes from 28 ARs are from Gopalswamy et al. (2017b); for one event, the RC flux was computed using SDO's HMI and AIA data. (a) is an updated version from Gopalswamy et al. (2010).

Figure 7 shows scatter plots of the magnetic potential energy of a large number of active regions with maximum speed and maximum kinetic energy of CMEs originating from the active regions. The active regions were selected based on the fact that they were responsible for one or more of the following: (i) a large SEP event, (ii) a magnetic cloud, and (iii) a major geomagnetic storm. The active region area was computed as the area covered by at least 10% of the peak unsigned magnetic field strength in the active region as observed in a magnetogram from SOHO/MDI or SDO/HMI when the region was close to the central meridian. The magnetic potential energy (MPE) is computed as $(<B>^2/8\pi)A^{1.5}$, where $<B>$ is the unsigned average field strength within $A$ (Gopalswamy et al. 2010). Note that the active region area used here is different from the sunspot area used in Fig. 6, which is typically smaller by a factor <10. On the other hand we use the average $B$, instead of the peak $B$ used for sunspots. In identifying the maximum CME speed in an active region, we listed out all the CMEs from the active region and selected the one with the highest speed to use in the scatter plot. The kinetic energy of the fastest CME from a given active region is taken from the SOHO/LASCO CME catalog. The MPE was computed at the time of the central meridian passage of an active region, and not at the time of the CME with the maximum speed.

Although the scatter is large, the maximum CME speed is significantly correlated with the MPE in the source active region (Fig. 7a), as was shown earlier in Gopalswamy et al. (2010). The regression line, when extrapolated to the maximum possible MPE ($3.7 \times 10^{36}$ erg), gives a speed of ~3600 km/s. This speed is not too different from the highest observed speed by SOHO/LASCO (see Fig. 3). Recall that the Weibull distribution gives this speed for a 100-year event (see Fig. 4). A straight-line fit to the top speeds in the plot corresponding to various MPEs would put the highest speed attainable by a CME as ~6700 km/s in an active region with a potential energy of ~$3.7 \times 10^{36}$ erg. This speed is greater than that of the 1000-year event (~4700 km/s) indicated by the Weibull distribution and similar to the speed indicated by the power-law distribution (~6500 km/s).

The maximum CME kinetic energy is also significantly correlated with MPE (see Fig. 7b). Not all CMEs had mass estimates, so the number of CMEs in the kinetic energy plot is smaller than that in the speed plot. The regression line (KE = $0.04$MPE$^{1.10}$) gives a kinetic energy of ~$4.2 \times 10^{35}$ erg corresponding to the highest possible MPE. This corresponds to an efficiency of energy conversion as ~11%. At lower levels of potential energy, the conversion efficiency is <10%. In the kinetic energy scatter plot, we have also shown the equal-energies line (100% efficiency). We do see a couple of data points that are close to the equal-energies line, but this may be due to the overestimate of the kinetic energies stemming from the uncertainties in mass and speed estimates.

Another way of looking at the energy conversion efficiency is to compare the reconnected (RC) flux during an eruption with the total active region flux. The total active region flux ($<B>A$) uses the same average magnetic field and area used in computing MPE ($<B>^2/8\pi)A^{1.5}$. The RC flux $\Phi_{RC}$ is computed as half the photospheric flux within the area under the post eruption arcade (Gopalswamy et al. 2017a). For a hypothetical region with the largest observed area (6100 msh) and the highest observed field strength (6000 G), the AR flux $\Phi_{AR}$ is ~$1.12\times10^{24}$ Mx. Substituting this value into the regression line ($\Phi_{RC} = 0.79\Phi_{AR}^{0.98}$), we get $\Phi_{RC}$ ~$2.9\times10^{23}$ Mx, suggesting that about 26% of the AR flux becoming reconnected in the eruption. Gopalswamy et al. (2017b) also reported an empirical relation between $\Phi_{RC}$ (in units of $10^{21}$ Mx) and the CME kinetic energy (in units of $10^{21}$ erg):

$$KE = 0.19(\Phi_{RC})^{1.87} \quad (2)$$

For $\Phi_{RC} = 2.9\times10^{23}$ Mx, this relation gives KE = $7.7\times10^{34}$ erg. This value is smaller by a factor of 5.5 than that ($4.2\times10^{35}$ erg) derived from the scatter plot in Fig. 7b. This is understandable because the KE in Fig. 7b is the maximum value for a give active region, while the one in equation (2) has no such constraint; it is simply computed for each eruption considered. The power law function (Y = 51.32 -1.58X where Y is the log of the occurrence rate per year and X is the KE) in Fig. 4 shows that KE = $7.7\times10^{34}$ erg gives an occurrence rate of $1.58\times10^{-4}$ per year; a CME with such KE will occur only once in ~6300 years.

In the above discussion we tacitly assumed that the free energy in active regions is released in the form of CME kinetic energy (eruptive flares). However, there may be no energy going into mass motion in the cases of confined flares. About 10% of X-class flares are known to be confined and the maximum size of a confined flare is ~X1.2 (Gopalswamy et al. 2009b). During solar cycle 24, a huge active region rotated from the east to the west limb of the Sun producing many major X-ray flares including an X3.1 flare. Although there were some narrow CMEs temporally coincided with a couple of the X-class flares, there was no CME associated with most of the X-class flares, including an X2 flare. The active region was NOAA 12192 with an area even larger than that of AR 10486 that resulted in the extreme space weather events shown in Fig.1. Even the change in the active region area was similar in the two regions (see Fig. 8). The magnetic potential energy of AR 12192 ($2.9\times10^{34}$ erg) was higher than that of AR10486 ($1.55\times10^{34}$ erg) by a factor of almost 2, but none of it went into mass motion. For such high magnetic potential energy, one would expect a CME with speed exceeding 3000 km/s from the correlation plot in Fig. 7b. Based on the investigation of the magnetic environment of AR 12192 it was concluded that the overlying field in the corona was so strong that it did not allow any mass to escape (Thalmann et al., 2015, and references therein). On the contrary, AR 10486 did not have the strong overlying field and had some connectivity to another active region nearby (AR 10484). Thus AR 12192 represents an extreme case in not producing any mass motion.

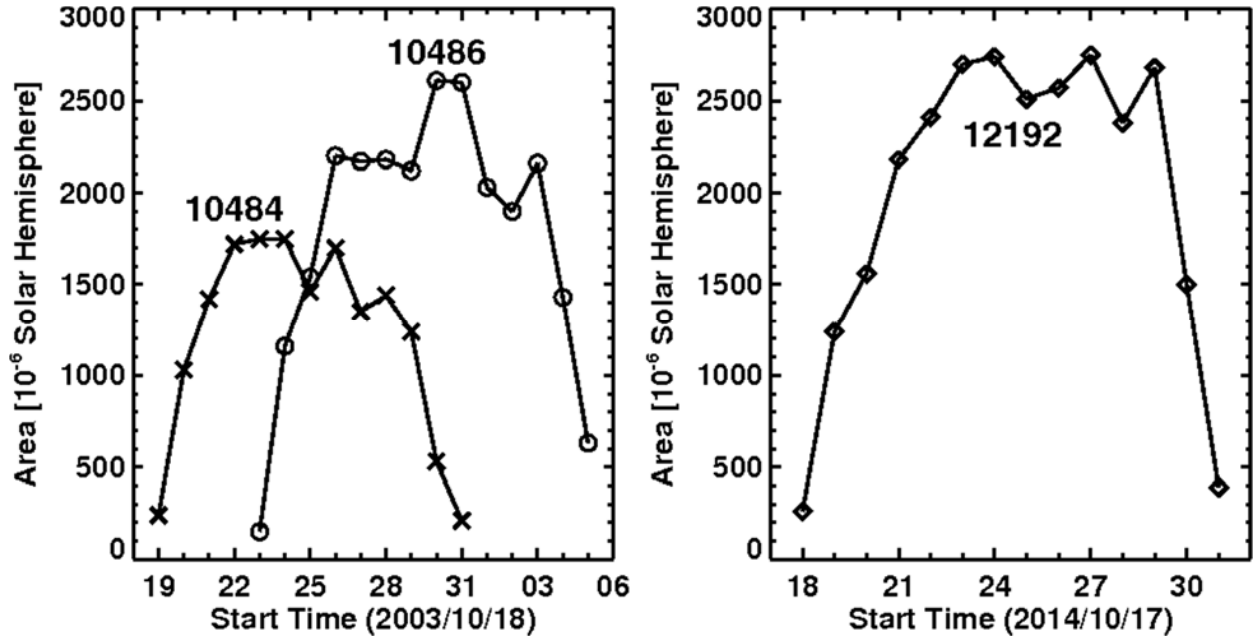

Figure 8. Observed time variation of the areas of two large active regions from (left) solar cycle 23 (October 2003, AR 10486) and (right) solar cycle 24 (October 2014, AR 12192). In the left plot, a nearby active region (AR 10484) with overlapping disk passage is also shown. The two active regions are at the extreme ends of eruptive behavior. AR data are from NOAA (http://www.swpc.noaa.gov/products/solar-region-summary).

## 4. Consequences of Solar Eruptions

The two primary consequences of CMEs are the SEP events and geomagnetic storms, of which the latter is specific to Earth. SEP events are relevant to any location in the heliosphere. In this section we consider the distributions of large SEP events (>10 MeV peak intensity ≥10 pfu) and intense geomagnetic storms (Dst ≤-100 nT). For SEP events, we also consider omnidirectional fluences in the >10 MeV and >30 MeV integral channels. We also discuss the tail of the distributions and how some of the historical events are located on the tails.

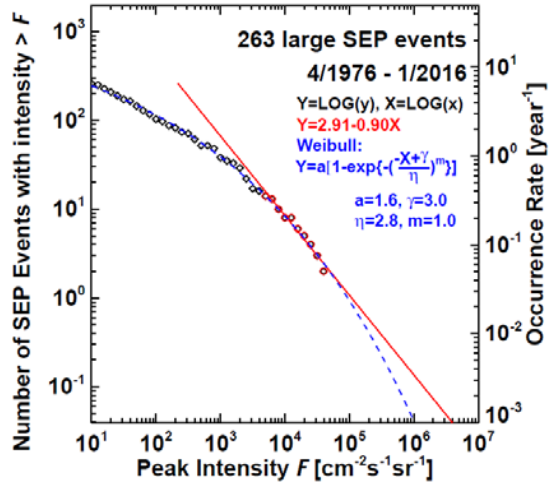

Figure 9. Cumulative distribution of large SEP events from 1976 to 2016 as reported by NOAA (also listed at NASA's Solar Data Analysis Center, https://umbra.nascom.nasa.gov/SEP/). A Weibull and power-law fits are shown. The power-law fitted only to the last 10 data points, whereas all data points are used in the case of Weibull distribution. The 23 March 1991 SEP event has the highest peak intensity of $4.3 \times 10^4$ cm$^{-2}$ s$^{-1}$ sr$^{-1}$.

## 4.1 SEP Events

Figure 9 shows the cumulative distribution of 261 large SEP events from 1976 to 2016 as reported by NOAA (https://umbra.nascom.nasa.gov/SEP/). All SEP events whose >10 MeV proton intensity exceeded anywhere during the event duration are included in the plot. This means, the largest events are energetic storm particle (ESP) events caused when the shock passes by the detector (see e.g., Cohen et al. 2006). The largest event has a size of ~$4.3 \times 10^4$ pfu, which is an ESP event that occurred on 1991 March 23 (Shea and Smart, 1993). The backside event of 2013 July 23 had a peak intensity of ~$4.4 \times 10^4$ pfu (Gopalswamy et al. 2016; Mewaldt et al. 2013), but it was a small event at Earth (~12 pfu). The Weibull fit can be extrapolated to obtain the size of 100-year and 1000-year events as $2.04 \times 10^5$ pfu and $1.02 \times 10^6$ pfu, respectively. The power-law fit gives even bigger sizes: $3.03 \times 10^5$ pfu and $3.96 \times 10^6$ pfu. It must be noted that both the power-law and Weibull fits do not pass through the last data point. If the largest measured value is correct, the extrapolated values may be overestimates. We can use the Ellison and Ramaty (1985) or Band et al. (1993) functions to force the fits pass through the last data point. The Band function gives the size of 100-year and 1000-year events as $9.51 \times 10^4$ pfu and $3.15 \times 10^5$ pfu, respectively. The Ellison-Ramaty function gives slightly lower values: $8.52 \times 10^4$ pfu and $1.57 \times 10^5$ pfu. We conclude that the maximum size of the 100-year event is ~$10^5$ pfu, while the size of the 1000-year can be an order of magnitude larger than this value.

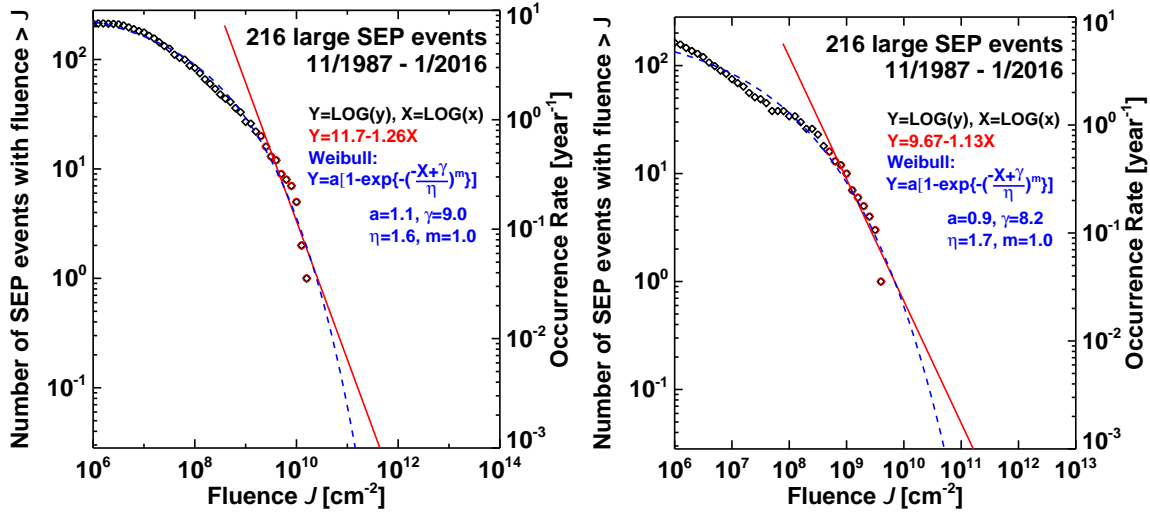

Figure 10. Cumulative distribution of the omnidirectional SEP fluence in the >10 MeV (a) and >30 MeV (b) ranges. Weibull and power-law fits are shown on the plots. The 14 July 2000 SEP event had the highest fluence of $1.65 \times 10^{10}$ cm$^{-2}$ (>10 MeV) and $4.31 \times 10^9$ cm$^{-2}$ (>30 MeV). All fluences were computed from time profiles of NOAA's GOES data.

## 4.2 SEP Fluences

Figure 10 shows the >10 MeV and >30 MeV fluences of 216 large SEP events detected by GOES since 1987. We have shown the Weibull and power-law fits to the occurrence rates. As in the intensity plot, the fitted curves do not pass through the last data point. Ellison-Ramaty (ER) and Band functions can be forced to pass through the last data point. The resulting 100-year and 1000-year fluences are compared in Table 1. The 100-year, >10 MeV fluence values are: $5.11 \times 10^{10}$ p cm$^{-2}$ (Weibull), $2.43 \times 10^{10}$ p cm$^{-2}$ (Ellison-Ramaty), and $2.48 \times 10^{10}$ p cm$^{-2}$ (Band). These values differ only by a factor of ~2. The 1000-year, >10 MeV fluence values are: $14.3 \times 10^{10}$ p cm$^{-2}$ (Weibull), $3.83 \times 10^{10}$ p cm$^{-2}$ (Ellison-Ramaty), and $4.94 \times 10^{10}$ p cm$^{-2}$ (Band). The Ellison-Ramaty and Band values are closer to each other, but the Weibull values are higher by a factor of 3-4. The 100-year fluence values for >30 MeV are: $1.58 \times 10^{10}$ p cm$^{-2}$ (Weibull), $0.63 \times 10^{10}$ p cm$^{-2}$ (Ellison-Ramaty), and $0.67 \times 10^{10}$ p cm$^{-2}$ (Band), while the 1000-year fluence values: $5.09 \times 10^{10}$ p cm$^{-2}$ (Weibull), $1.02 \times 10^{10}$ p cm$^{-2}$ (Ellison-Ramaty), and $1.52 \times 10^{10}$ p cm$^{-2}$ (Band). The Ellison-Ramaty and Band values are consistently close to each other, while the Weibull values are larger by a factor 3-5. The power law fits yield higher values in all cases, by about an order of magnitude.

Table 1. Integral fluence values for different models in units of $10^{10}$ p cm$^{-2}$

| Model | 100-year | | 1000-year | |
|---|---|---|---|---|
| | >10 MeV | >30 MeV | >10 MeV | >30 MeV |
| Weibull | 5.11 | 1.58 | 14.3 | 5.09 |
| Power-law | 7.08 | 2.12 | 43.7 | 16.3 |
| Ellison-Ramaty | 2.43 | 0.63 | 3.83 | 1.02 |
| Band | 2.48 | 0.67 | 4.94 | 1.52 |

Based on SEP event identification made from nitrate deposits in polar ice, Shea et al. (2006) compiled the >30 MeV fluences of events that occurred over the past ~450 years. They concluded from the frequency distribution of these events that the occurrence of > 30 MeV fluence exceeding $0.6 \times 10^{10}$ p cm$^{-2}$ are very rare. However, Wolff et al. (2012) have questioned the statistics on the basis of their finding that most of the nitrate spikes in Greenland ice cores correspond to biomass burning plumes originating in North America. They were also not able to find a nitrate signal even for the Carrington event. In fact, in a simulation study, Duderstadt et al. (2016) concluded that an SEP event large enough and hard enough to produce a nitrate signal in Greenland ice core would not have occurred throughout the Holocene. This conclusion is consistent with the > 30 MeV, 100-year fluences obtained in this study. The estimated largest, >30 MeV fluence of $0.6 \times 10^{10}$ p cm$^{-2}$ was also reported by Webber et al. (2007) for the 1960 November 12 GLE event. Cliver and Dietrich (2013) also estimated the >30 MeV integral fluence to be in the range $(0.5\text{-}0.7) \times 10^{10}$ p cm$^{-2}$ for a few GLE events (1959 July, 1960 November, and 1972 August). Their highest estimate was for the Carrington event: $1.1 \times 10^{10}$ p cm$^{-2}$ similar to our 100-year fluence from the Weibull distribution. Cliver and Dietrich (2013) noted that the Carrington event is a composite event due to multiple eruptions that happened in quick succession.

Extending the historical data over longer periods became possible with the discovery of two possible SEP events in tree rings. Measurements of $^{14}$C in Japanese cedar trees revealed significant increases in the carbon content during two periods: AD774–775 (Miyake et al. 2012) and AD 992-993 (Miyake et al. 2013). The authors concluded that the two events must be of similar origin. The two events were also identified in Antarctic and Arctic ice core as enhancements in cosmogenic isotopes such as $^{10}$Be and $^{36}$Cl (Mekhaldi et al. 2015). There has been considerable debate on the origin of these events (Melott and Thomas, 2012; Usoskin et al. 2013; Hambaryan and Neuhäuser 2013; Pavlov et al. 2013; Cliver et al. 2014), but the idea that these are due to SEP events seems to be gaining acceptance (Mekhaldi et al. 2015; Usoskin 2017). In particular, Mekhaldi et al. (2015) provided arguments against cometary and gamma ray burst sources. They also confirmed that a SEP event with a hard spectrum above 100 MeV is needed to cause these enhancements, as suggested by Usoskin et al. (2013). For the present discussion, we take the AD774/5 and AD 992/3 signals to be consequences of SEP events and compare them with the fluences we obtained in Table 1.

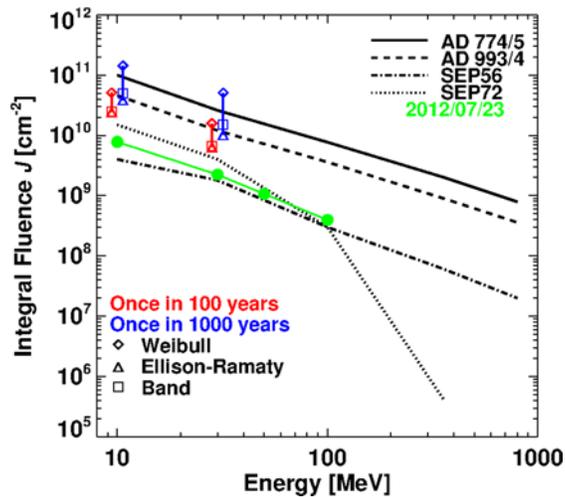

Figure 11. The 100-year and 1000-year data points derived from the cumulative distributions are superposed on the spectra of the AD774 and AD 993 particle events obtained by Mekhaldi et al. (2015). Estimates of 100-year and 1000-year event sizes from Weibull, Ellison-Ramaty, and Band functions are shown using different symbols. The data points are shifted slightly to the left (100-year) and right (1000-year) of X=10 MeV and X=30 MeV to distinguish them. The spectra of the 1956 February 23 (SEP56) and 1972 August 4 (SEP72) solar proton events are also shown from Mekhaldi et al. (2015), who used the reevaluated spectra from Webber et al. (2007). Also shown is the spectrum of the 2012 July 23 extreme event from Gopalswamy et al. (2016).

Figure 11 shows the estimated fluence spectra of the AD774/5 and AD 992/3 events from Mekhaldi et al. (2015) obtained by scaling the hard spectrum of the 2005 January 20 GLE event. Also shown for comparison are the hard spectrum of the 1956 February 23 GLE, the soft spectrum of the 1972 August 4 GLE, and a recent event on 2012 July 23, which most likely accelerated particles to GeV energies. Superposed on these plots are the 100-year and 1000-year fluences obtained from Fig. 10 using Weibull, Ellison-Ramaty, and Band functions. Clearly, the 100-year and 1000-year fluences are consistent with those of the AD774/5 and AD 992/3 events. In particular, the 1000-year fluences in the >10 MeV and >30 MeV ranges cover the AD774/5 and AD 992/3 events with the two-point slope consistent with that of the known SEP events. This comparison also supports the possibility that the AD774/5 and AD 992/3 events are indeed consequences of SEP events.

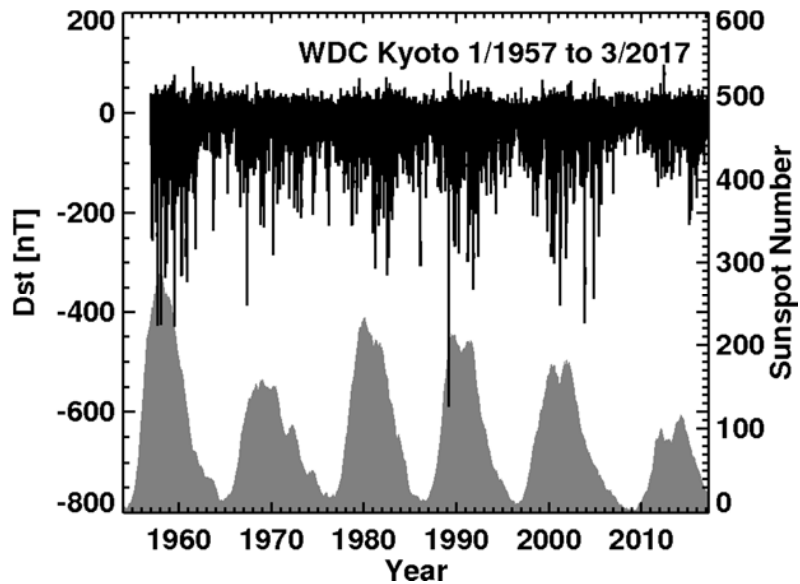

Figure 12. A plot of the Dst index available at the World Data Center (WDC) in Kyoto, Japan from 1957. The large negative excursions below -100 nT are major storms. The sunspot number is shown at the bottom (gray) for reference. The largest storm occurred on 1989 March 13.

### 4.3 Large Geomagnetic storms

Geomagnetic disturbances have been recognized since the 1600s and the term geomagnetic storm was introduced by von Humboldt in the 1800s (see Howard 2006 for a review). The link to the Sun was recognized by Sabine (1852) as a synchronous variation of sunspot number and geomagnetic activity. The Carrington flare occurred a few years later and was associated with a geomagnetic storm of historical proportions. Fortunately, there were extensive observations of the storm from magnetometers and global aurora (e.g., Tsurutani et al. 2003). This remains a historical extreme event against which many storms are compared (Cliver and Dietrich 2013). The connection between solar eruptions and geomagnetic storms was established in the early 20[th] century including the fact that the eruptions occurred close to the disk center of the Sun and an average delay of ~1 day was noted between the flare occurrence and the onset of great geomagnetic storms (Hale 1931; Newton 1943).

Now we know that the magnetic field in CMEs and in the sheath ahead of shock-riving CMEs is responsible for intense geomagnetic storms (Wilson 1987; Gonzalez and Tsurutani 1997). In particular, the strength of the CME magnetic field component oriented in the direction opposite to that of Earth's horizontal magnetic field Bz is critically important in causing intense storms along with the speed (V) with which the CME magnetic field impinges on the magnetosphere (see e.g., Wu and Lepping 2002; Gopalswamy et al. 2008). The following empirical relation reasonably represents how the storm strength (Dst) is determined by V and Bz (Gopalswamy 2010b):

$$Dst = -0.01V|Bz| - 32 \text{ nT.} \qquad (3)$$

The Dst index has been compiled since 1957 and has identified many modern super magnetic storms. Figure. 12 shows a plot of the Dst index as a function of time along with the sunspot number. There are only five super storms that had Dst index < -400 nT. The 1989 March 13 storm is the largest since 1957 with a Dst of -589 nT. The storm was associated with a series of solar eruptions between March 10 and 12. The primary storm started with a sudden commencement at 07:47 UT on March 13. The storm has been attributed to a solar eruption, which occurred at 00:16 UT on 1989 March 12 from N28E09. Details on the solar source (Zhang et al. 1995) and the interplanetary conditions (Nagatsuma et al. 2015) have been reported before. Even though the storm was an extreme event, the flare itself was of moderate size (M7.3). There was no CME data for this event, but from the transit time of 31.5 h one can infer that the CME had a transit speed of ~ 1300 km/s. Direct solar wind measurements were also not available, so one has to infer the speed based on empirical relations (Cliver at al. 1990; Belov et al. 2008) as ~960 km/s. From equation (3) with Dst = -589 nT, we see that $VBz = 6.2 \times 10^4$ nT.km s$^{-1}$. For V = 960 km/s, we get Bz = -65 nT. Nagatsuma et al. (2015) estimated a Bz of ~ -50 nT.

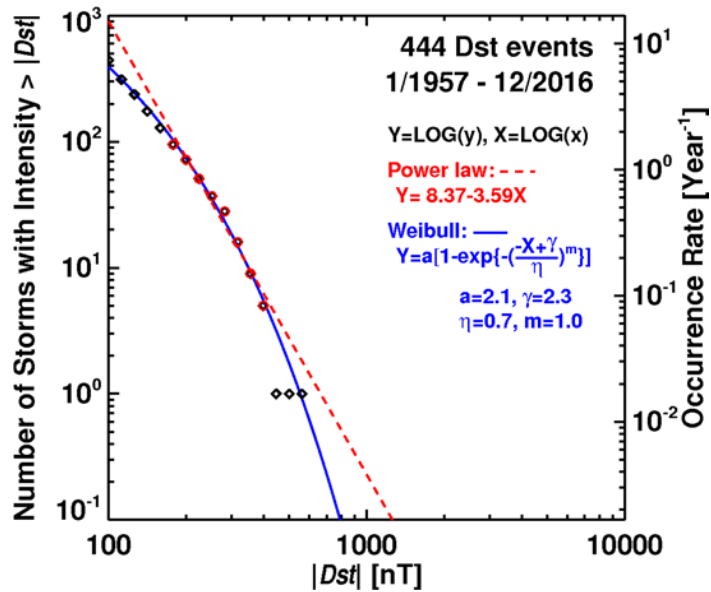

Figure 13. Cumulative distribution of intense geomagnetic storms (Dst ≤-100 nT) and their yearly rates using Dst data from 1957 available at the World Data Center, Kyoto. Weibull and Power-law fits to the distribution are shown. The largest storm (|Dst| = -589 nT) occurred on 1989 March 14 at 02:00 UT.

Figure 13 shows the cumulative distribution of intense geomagnetic storms from the Dst data made available on line at the Kyoto World Data Center. As in many other distributions, the power law fit seems to overestimate the 100-year and 1000-year events. The Weibull distribution fits all the data points. According to the Weibull distribution, a 100-year event has a size of -603 nT, consistent with the March 1989 event; a 1000-year event has a size of -845 nT, consistent with some estimates of the Carrington storm, which occurred about 157 years ago. Although the Dst equivalent of the Carrington storm was estimated as -1600 nT from the geomagnetic record at the Colaba Observatory in India (Tsurutani et al. 2003), many authors have argued for a downward revision. The main arguments are: (i) the Dst index is an hourly average and (ii) ionospheric/auroral currents might have contributed to the initial sharp spike recorded at the Colaba observatory (see Cliver and Dietrich 2013 for details). Applying hourly averages to the Colaba data, Siscoe (2006) arrived at a Dst index of -850 nT, similar to the 1000-year event from the Weibull distribution. Recently, Gonzalez et al. (2011) reanalyzed that Colaba data and arrived at a Dst equivalent of ~1160 nT. It must be noted that these estimates are also approximate because the Dst index is actually an average over several equatorial magnetic observatories (see http://wdc.kugi.kyoto-u.ac.jp/dstdir/dst2/onDstindex.html). Cliver and Dietrich suggest a Dst of -900 nT as a nominal value for the Carrington event.

The shock transit time of 2381 km/s when used in Cliver et al (1990) empirical relation,

$$V = 0.775Vt - 40 \text{ km/s} \quad (4)$$

gives a shock speed at Earth of ~1800 km/s. The CME speed near the Sun is expected to be ~3000 km/s. This speed is twice that of the 1989 March 13 event, and hence a doubling of the Dst index is not unexpected. We can also get the 1-AU shock speed of the Carrington event from the average acceleration ($a$) reported in Gopalswamy et al. (2001):

$$a = -0.0054 \ (u - 406) \text{ m/s}^2, \quad (5)$$

where $u$ is the CME speed near the Sun. A lower limit to $a$ can be obtained by replacing the initial speed by the transit speed in equation (5), yielding $a = -10.7 \text{ ms}^{-2}$. From the transit speed and the deceleration, we get a 1-AU speed of ~2044 km/s, only13% larger than the speed from eq. (4). The initial CME speed can then be estimated as 2700 km/s. Such a speed is well within the observed range of CMEs (see Fig. 3). For Dst = -900 nT and V=2044 km/s, Bz can be estimated as ~-46 nT. For Dst = -1160 nT, only a Bz of -58 nT is needed. These estimates are reasonable if the storm was caused by the shock sheath. If the storm was due to the ICME, one has to allow for the possibility of an ICME speed (V$_{ICME}$) lower than the shock speed. Using the gas dynamic strong shock limit,

$$V = V_{ICME} \ (1+\gamma)/2 \quad (6)$$

where $\gamma$ is the adiabatic index. For $\gamma$=5/3 and V = 2044 km/s, eq. (6) gives V$_{ICME}$ = 1533 km/s. In this case, Dst = -900 nT and -1160 nT would require a Bz of – 61 nT and -78 nT, respectively. These numbers are consistent with a recent backside CME on 2012 July 23 that had Bz ~ -52 nT, V ~2000 km/s, and V$_{ICME}$ ~ 1560 km/s at 1-AU (Gopalswamy et al. 2016). The storm strength has been estimated to be similar to that of the Carrington event (Baker et al. 2013; Russell et al. 2013; Mewaldt et al. 2013; Liu et al. 2014; Gopalswamy et al. 2015a).

The requirement of Bz = -78 nT to get a Dst value of -1160 nT is not unlikely. Gopalswamy et al. (2017b) obtained an empirical relationship between the peak total magnetic field strength (Bt) and ICME speed:

Bt = 0.06 V$_{ICME}$ - 13.58 nT (7)

For V$_{ICME}$ =2000 km/s, eq. (7) can be extrapolated to give Bt = 106 nT. From the compilation of Bz and Bt for cycle-23 magnetic clouds in Gopalswamy et al. (2015a), we can see that the magnitude of Bz ~0.74Bt, thus yielding Bz ~ -78 nT for Bt = 106 nT. Thus we conclude that the Carrington storm can be explained by a very fast ICME with high magnetic content and the Dst estimate is consistent with a 1000-year storm.

Table 2. Expected 100-year and 1000-year event sizes estimated from the tail of observed distributions fitted to various functions.

| | 100-year | | 1000-year | |
|---|---|---|---|---|
| | Weibull | Power law | Weibull | Power law |
| AR Area (msh) | 5780 | 7090 | 8200 | 13600 |
| CME speed (km/s) | 3800 | 4484 | 4670 | 6564 |
| CME KE ($10^{33}$ erg) | 4.40 | 6.85 | 9.76 | 29.5 |
| Flare Size (X1.0=$10^{-4}$W m$^{-2}$) | X43.9 | X42.4 | X101 | X115 |
| Bolometric Flare Energy ($10^{32}$ erg) | 4.39 | 4.24 | 10.1 | 11.5 |
| SEP Intensity ($10^5$ pfu) | 2.04 | 3.03 | 10.2 | 39.6 |
| >10 MeV SEP Fluence ($10^{10}$ cm$^{-2}$) | 5.11 | 7.07 | 14.3 | 43.7 |
| >30 MeV SEP Fluence ($10^{10}$ cm$^{-2}$) | 1.58 | 2.12 | 5.09 | 16.30 |
| Dst (nT) | -603 | -774 | -845 | -1470 |

## 5. Summary and Conclusions

This chapter considered properties of extreme solar eruptions and their consequences assuming that they are located on the tail of their cumulative distributions. In particular, we estimated the sizes of 100-year and 1000-year events. In many cases, these sizes are consistent with known historical events. Weibull unreliability function was used as the baseline function in extrapolating the distributions to estimate the 100-year and 1000-year event sizes. Power-law distributions were also used, but generally they appear to yield overestimates. In some cases, we also used Ellison-Ramaty and Band functions in obtaining conservative estimates of 100-year and 1000-year events. The power laws can be fit only to a subset of the data points and their selection is somewhat subjective. In some cases the even the Weibull distribution may lead to overestimates, but not by as large an extent. Table 2 provides a summary of the 100-year and 1000-year event sizes as a measure of the extremeness of the phenomena considered. The range of values for a given entry between the power law and Weibull distributions give an idea of the uncertainties involved in the event size estimations.

We also considered solar active regions as the physical origin of eruptive events and considered the maximum amount of free energy available for powering the eruptions. The limit to the free energy can be traced to the size and magnetic content of active regions. The free energy in an active region is generally not exhausted in a single eruption, so the maximum flare size or the CME kinetic energy is limited by a conversion efficiency, which is not fully understood. Two decades of SOHO observations have shown that there are not many CMEs with speeds exceeding 3000 km/s. The tail of the Weibull distribution suggests that a 1000-year CME will have a speed of only 4700 km/s. A 1000-year CME is expected have a kinetic energy of $\sim 10^{34}$ erg. Similarly, a 1000-year flare will have a size of $\sim$X100; the corresponding bolometric flare energy of $10^{33}$ erg is consistent with the known fact that the CME kinetic energy is typically ten times the flare energy.

The consequences of eruptive events occurs we considered are SEP events and geomagnetic storms. We estimate the >30 MeV fluence of a 1000-year event is in the range $(1-5) \times 10^{10}$ p cm$^{-2}$, which is consistent with the historical extreme event such as the Carrington event, the AD 774/75 event, the AD 994/95 event, and the recent backside event of 2012 July 23. The Carrington event also serves as the bench-mark geomagnetic storm. The tail of the Weibull distribution gives the Dst index of a 1000-year event as -845 nT, which is consistent with the revised estimates of the Carrington storm size. The power law tail gives a larger storm magnitude consistent with higher estimates for the Carrington event, although we think the power law overestimates the event sizes.

**Acknowledgments.**

I thank P. Mäkelä, S. Akiyama, and S. Yashiro for help with the figures. I thank D. F. Webb and E. W. Cliver for their comments and suggestions that improved the presentation of the material. This work was supported by NASA's Heliophysics Guest Investigator program.

Acronyms

AR – Active region

CMD – Central meridian distance

CME – Coronal mass ejection

Dst – Disturbance storm time (index)

EIT – Extreme-ultraviolet imaging telescope

ESP – Energetic storm particle

FOV – Field of view

GLE – Ground level enhancement

GOES – Geostationary Operational Environment Satellite

GPS – Global positioning system

HMI – Helioseismic and Magnetic Imager

KE – Kinetic Energy

LASCO – Large Angle and Spectrometric Coronagraph

MDI – Michelson Doppler Imager

MPE – Magnetic potential energy

MSH – millionths of solar hemisphere

SC – Sudden commencement of geomagnetic storms

SDO – Solar Dynamics Observatory

SEP – Solar energetic particle

SOHO – Solar and Heliospheric Observatory

SOON – Solar Observing Optical Network

STEREO – Solar Terrestrial Relations Observatory

WDC – World Data Center

**Biographical Sketch: Dr. Nat Gopalswamy, Astrophysicist, NASA Goddard Space Flight Center**

Dr. Nat Gopalswamy is an Astrophysicist with the Solar Physics Laboratory, Heliophysics Division of NASA's Goddard Space Flight Center. He is an internationally recognized expert in coronal mass ejections and their space weather consequences, with a deep interest in understanding how the solar variability impacts Earth. He has over 30 years of experience in solar-terrestrial research, working on projects such as SOHO, Wind, STEREO, and SDO. He is also a solar radio astronomer working on thermal and nonthermal radio emission from the Sun using data the Clark Lake Radioheliograph, the Very Large Array, and the Nobeyama Radioheliograph. He has authored or co-authored more than 400 scientific articles and has edited nine books. He has received numerous awards and honors including the 2013 NASA Leadership Medal. He is currently the President of ICSU's Scientific Committee on Solar Terrestrial Physics (SCOSTEP) and the Executive Director of the International Space Weather Initiative (ISWI). He is a Fellow of the American Geophysical Union.